\documentclass[aps,prl,twocolumn,showpacs,superscriptaddress,groupedaddress]{revtex4-1}

\usepackage{graphics, bm, xspace}
\usepackage{graphicx}
\usepackage{psfrag}
\usepackage{amsmath}
\usepackage{amssymb}
\usepackage{epsfig}
\usepackage{subfigure}
\usepackage{grffile}
\usepackage{times}
\usepackage{color}
\usepackage{multirow}
\usepackage[bookmarks=false,linkcolor=blue,urlcolor=blue,colorlinks,citecolor=blue]{hyperref}
\usepackage{float}
\usepackage{gensymb}

\newcommand{\beq}    {\begin{equation}}
\newcommand{\enq}    {\end{equation}}
\newcommand{\ceq}[1] {(\ref{#1})}

\newcommand{\eps}    {\epsilon}

\newcommand{\kk}     {{\bf k}}

\newcommand{\qq}     {{\bf q}}

\newcommand{\JJ}     {{\bf J}}

\newcommand{\cross}  {\times}

\newcommand{\df}     {\equiv}

\newcommand{\rr}     {{\bf r}}
\newcommand{\nimp}   {n_{\rm imp}}

\newcommand{\bise}        {${\rm Bi_2Se_3}$\xspace}

\newcommand{\ssigma}     {{\boldsymbol\sigma}}

\graphicspath{{../Figures/}}

\usepackage[latin1]{inputenc}
\usepackage{amsmath}
\usepackage{amsfonts}
\usepackage{amssymb}
\usepackage{graphicx}
\usepackage{float}
\usepackage{subfigure}
\usepackage{algorithm}
\usepackage{subfigure}
\usepackage{xcolor}
\usepackage{cancel}
\usepackage{bm}

\newcommand*\colvec[3][]{
    \begin{bmatrix}\ifx\relax#1\relax\else#1\\\fi#2\\#3\end{bmatrix}
}

\begin{document}

\title{Spin-charge coupled transport in van der Waals systems with random tunneling}
\author{M. Rodriguez-Vega$^{1,2}$, G. Schwiete$^3$, Enrico Rossi$^4$}
\affiliation{
             $^1$Department of Physics, The University of Texas at Austin, Austin, TX 78712, USA\\
             $^2$Department of Physics, Northeastern University, Boston, MA 02115, USA\\
             $^3$Department of Physics and Astronomy, Center for Materials for Information Technology (MINT), The University of Alabama, Alabama 35487, USA\\
             $^4$Department of Physics, William \& Mary, Williamsburg, VA 23187, USA
             }
\date{\today}
\begin{abstract}
We study the electron and spin transport in a van der Waals system formed by one layer with strong spin-orbit coupling and a second layer without spin-orbit coupling, in the regime when the interlayer tunneling is random. We find that in the layer without intrinsic spin-orbit coupling spin-charge coupled transport can be induced by two distinct mechanisms. First, the gapless diffusion modes of the two isolated layers hybridize in the presence of tunneling, which constitutes a source of spin-charge coupled transport in the second layer. Second, the random tunneling introduces spin-orbit coupling in the effective disorder-averaged single-particle Hamiltonian of the second layer. This results in non-trivial spin transport and, for sufficiently strong tunneling, in spin-charge coupling. As an example, we consider a van der Waals system formed by a two-dimensional electron gas (2DEG)--such as graphene--and the surface of a topological insulator (TI) and show that the proximity of the  TI induces a coupling of the spin and charge transport in the 2DEG. In addition, we show that such coupling can be tuned by varying the doping of the TI's surface. We then obtain, for a simple geometry, the current-induced non-equilibrium spin accumulation (Edelstein effect) caused in the 2DEG by the coupling of charge and spin transport.
\end{abstract}	
\maketitle

%
In recent years experimentalists have been able to make very novel and high quality heterostructures that allow
the realization of new effects and states of great fundamental and technological interest~\cite{geim2013}. Recently simple heterostructures formed by two graphene layers with a relative twist angle~\cite{dossantos2007,mele2010,bistritzer2011,lu2016}
have shown a phase diagram~\cite{cao2018,yankowitz2019} that is remarkably reminiscent of the phase diagram of high temperature superconductors.
These are just some of the most striking examples that heterostructures can be used to realize novel
effects that are not present in the single constituents. 
Applications of heterostructure engineering \cite{novoselov2016} can be found in tunnel junctions \cite{lee2011}, plasmonic \cite{woessner2014},  photoresponsive \cite{roy2013}, spintronics \cite{zutic2004,sinova2015,cardoso2018} and valleytronic \cite{schaibley2016} devices.

One of the essential elements to realize non-trivial topological states and spin-dependent transport phenomena is the presence
of spin-orbit coupling (SOC). However, often the presence of spin-orbit coupling is not accompanied by other desirable 
properties such as high mobility, or superconducting pairing. For this reason 
heterostructures that combine one constituent with significant SOC and one constituent with no SOC but other distinct properties
are very interesting both for fundamental reasons and for their potential for technological applications.
One example, of such heterostructures are  graphene--topological-insulator van der Waals 
systems~\cite{dang2010,kou2013,jin2013,jzhang2014,rodriguez2014, lzhang2016,
mrv2017,song2018,khokhriakov2018}.
So far, the theoretical studies of van der Waals heterostructures have focused on the regime 
when the tunneling is not random and a strong hybridization between the electronic states
of the isolated systems can be achieved. However, in many situations we can expect the tunneling between the systems
forming the heterostructure to be random, due for example to the incommensurate nature of the stacking configuration
and/or the presence of surface roughness. 

In this work we focus on this situation, and study the electron and spin transport in
a two-dimensional van der Waals systems comprised of one component (layer) with strong SOC and one with no, or negligible, SOC, 
when the interlayer tunneling is random.
Due to the random nature of the tunneling in most experimental situations the transport will be 
diffusive even in the absence of disorder. For this reason we consider only the diffusive regime, in which specific details of the system considered (like the value of the mean free path)
do not affect the general expression of the transport equations that, therefore, have a somewhat universal character.
We find that in general, if the diffusive transport in the layer with SOC exhibit 
spin-charge coupling~\cite{burkov2004,burkov2010,shen2014}
such coupling will be present also in the layer without SOC, i.e., in the most common experimental situation.
To exemplify this general result we consider the case of a van der Waals system formed
by a two-dimensional electron gas (2DEG) placed on the surface of a strong three dimensional topological insulator (TI)~\cite{hasan2010,qi2011}.
Graphene and the surface of TIs in the tetradymite family such as \bise have almost commensurate lattices and as
a consequence in many graphene-TI heterostructures the $K$, $K'$ points of the graphene's BZ are folded 
close to the TI's $\Gamma$~\cite{jzhang2014} point. This fact, combined with the random and finite-range nature
of the interlayer tunneling, implies that the results that we obtain for a 2DEG-TI van der Waals system 
are directly relevant to graphene-TI heterostructures, and similar systems.
We obtain the diffusive transport equations in the 2DEG layer and show that they describe 
a transport in which the charge and the spin degrees of freedom are coupled. Finally, we show how the diffusive equations give rise to spin-dependent transport effects,
analogous to the ones obtained for a 2DEG with Rashba SOC~\cite{burkov2004} and an isolated TI's surface~\cite{burkov2010},
that are tunable by simply varying the doping of the TI, and that can be used for possible spintronics applications.


The Hamiltonian $\hat H$ for the heterostructure can be written as 
$\hat H = \sum_{l=1,2}[ \hat H_{l} + \hat V_l]+  \hat T$
where $l$ is the layer index, $\hat H_l$ is the Hamiltonian for layer $l$ in the clean limit,
$\hat V_l$ is the term due to disorder located in layer $l$, and $\hat T$ is the term describing interlayer tunneling. For the 2DEG layer we have
$ \hat H_l = \hat H_{\rm 2d}(\kk) = \sum_{\kk ss'} \hat \psi^\dagger_{{\rm 2d}, \kk s}  H_{{\rm 2d} ss'}(\kk) \hat \psi_{{\rm 2d}, \kk s'}$ 
where, 
$\hat \psi^\dagger_{\rm 2d, \kk s} $  $( \hat \psi_{{\rm 2d}, \kk s} )$ is the creation (annihilation) operator for
an electron with momentum $\kk$ and spin $s$.
Without loss of generality we can linearize the 2DEG dispersion around the Fermi surface and assume
$H_{{\rm 2d} }(\kk)= (v_{\rm 2d} |\kk| -\mu_{\rm 2d}) \sigma_0$
with 
$v_{\rm 2d}$ the Fermi velocity, $\mu_{\rm 2d}$ the chemical potential, and
$\sigma_0$ the $2\times 2$ identity Pauli matrix in spin space. For the TI's surface we have
$ \hat H_l =\hat H_{\rm TI} = \sum_{\kk ss'} \hat \psi^\dagger_{{\rm TI}, \kk s} H_{{\rm TI} ss'}(\kk) \hat \psi_{{\rm TI}, \kk s'}$
where 
$\psi^\dagger_{{\rm TI}, \kk s}$ $(\psi^\dagger_{{\rm TI}, \kk s})$ creates (annihilates) a surface Dirac fermion with spin $s$ and momentum $\kk$, $ H_{\rm TI}(\kk) = - v_{\rm TI} \left( \kk \times \bm{\sigma} \right)_z-\mu_{\rm TI}$, $v_{\rm TI}$ being the Fermi velocity on the TI's surface, $\mu_{\rm TI}$ the TI's surface chemical potential,
and $\sigma_i$, $i=x,y$ the Pauli matrices in spin space.

For the disorder potential in layer $l$, $V_l^{(D)}(\qq)$, we have 
$\langle V_l^{(D)}(\rr_1) V_{l}^{(D)}(\rr_2) \rangle = W_l^D(\rr_1 - \rr_2)$,
where the angle brackets denote average over disorder realizations,
and $W^D_l(\rr_1-\rr_2)$ is the disorder-averaged spatial correlation.
In momentum space we have $W_l^D(\qq)=\nimp^l |U(\qq)|^2$ where $\nimp^l$ is the impurity density in layer $l$,
and $U_l(\qq)$ the Fourier transform of the potential profile $U_l(\rr)$ of a single impurity.
Without loss of generality we can assume $\langle  V^{(D)}_l(\rr) \rangle =0$. Assuming the tunneling to be spin-conserving we have $\hat T = \sum_{\kk \qq s} T(\qq) \hat \psi^\dagger_{\bar l \kk s} \hat \psi_{l \kk+\qq s} + h.c.$ with $\bar l \neq l$.
Assuming the tunneling to be random we can characterize it by the spatial average of the tunneling matrix element $\langle T({\bf r_1})T({\bf r_2}')\rangle=W^t({\bf r_1}-{\bf r_2})$. In the remainder
we assume both the intralayer disorder and interlayer tunneling to be short-range 
so that $U_l(\qq)={\rm const}=U_l$, $W^t(\qq)={\rm const}=t^2$.

Let 
$G_{ 0 l }^{R,A}(\kk, \epsilon) = \left( \epsilon - H_{l}(\kk) \pm 0^+\right)^{-1}$
be the bare retarded (advanced) real-time Green's function for layer $l$. The total self-energy for layer $l$,  $\Sigma_l$, has contributions from scattering with impurities, $\Sigma^0_{l}$, 
and random tunneling events $\Sigma^t_{l}$. We have $\Sigma^0_{l}(\kk,\epsilon)= n^l_{imp} \int_\qq |U_l(\qq)|^2  G_l(\kk-\qq,\epsilon)$, 
where $\int_\qq\equiv\int {d^2 \qq}/{(2\pi)^2}$. 
In the self-consistent Born approximation, $G_l$ is the disorder-dressed Green's function for layer $l$.
For the 2DEG, apart from an overall unimportant real constant,
we have
$\Sigma^0_{2d}=-i\Gamma^0_{2d}\sigma_0/2$,
where 
$\Gamma^0_{2d}=1/\tau^0_{2d}=2\pi\rho_{2d} n^{2d}_{imp} U_{2d}^2$,
and $\rho_{2d}$
is the density of states (DOS) at the Fermi energy. For the TI's surface, due to the fact that the electrons behave as massless Dirac fermions, for  $U_{\rm TI}(\qq)={\rm const}$,
we have that the integral in the expression for $\Sigma^0_l$
has an ultraviolet divergence~\cite{sakai2014}.
After properly regularizing such divergence~\cite{fujimoto2013} one finds that
the intralayer disorder, in addition to generating an imaginary part of the self-energy,  
$-i\Gamma^0_{\rm TI}\sigma_0$, with 
$\Gamma^0_{\rm TI}=1/\tau^0_{\rm TI}=\pi\rho_{\rm TI} n^{\rm TI}_{imp} U_{\rm TI}^2$ and $\rho_{\rm TI}$,
the TI's DOS at the Fermi energy, causes a renormalization of the Fermi velocity that we incorporate in the definition of $v_{\rm TI}$. The same ultraviolet divergence appears for the self-energy 
correction for the 2DEG due to tunneling events into the TI, $\Sigma^t_{\rm 2d}$.
The proper renormalization of such divergence, consistent with the Ward identities, 
causes $\Sigma^t_{\rm 2d}$ to have a non-trivial real part so that
\beq
 \Sigma^t_{\rm 2d}(\kk,\epsilon) = -i\Gamma^t_{\rm 2d}\sigma_0/2 + \left(t^2/(4\pi v_{\rm TI}^2)\right) \left(\kk\cross\ssigma \right)_z.
 \label{eq:sigmat_TI}
\enq
where $\Gamma^t_{\rm 2d}=1/\tau^t_{\rm 2d}=\pi\rho_{\rm TI}t^2$.
This result shows that even when the interlayer tunneling processes are random,
a spin-orbit coupling term is induced in the 2DEG due to TI's surface proximity.
This term of the self energy qualitatively affects the diffusive transport in the 2DEG, but {\em it is not necessary}
to induce spin-charge transport in the 2DEG as we will show below. The  self-energy correction for the TI due to tunneling events into the 2DEG, $\Sigma^t_{\rm TI}$, does not require any special
care and simply results in an additional broadening of the quasiparticles: $\Sigma^t_{\rm TI}(\kk,\eps)=-i\Gamma^t_{\rm TI}\sigma_0/2$ with $\Gamma^t_{\rm TI}=1/\tau^t_{\rm TI}=2\pi\rho_{\rm 2d}t^2$.

With the self-energy contributions, the dressed 2D system Green's functions take the form
\begin{align}
 G_{{\rm 2d},\epsilon}^{R/A}(\kk) &= \frac{(\epsilon\pm i \Gamma_{2d}/2 -\epsilon_{\kk})\sigma_0 - [t^2/(4\pi v_{\rm TI})] \left( \kk \times \sigma \right)_{z}}
                                   {(\epsilon\pm i\Gamma_{\rm 2d}/2-\epsilon_\kk)^2 - [t^4/(4\pi v_{\rm TI})^2]k^2} \;, \label{eq:G2d}\\
 G_{{\rm TI},\epsilon}^{R/A}(\kk) &= \frac{(\epsilon\pm i\Gamma_{\rm TI}/2)\sigma_0 -  v_{\rm TI} \left( \kk \times \sigma \right)_{z}}
                                   {(\epsilon\pm i\Gamma_{\rm TI}/2)^2 -  v^2_{\rm TI} k^2}, \label{eq:GTI}
\end{align}
where $\Gamma_{\rm 2d}\df \Gamma^0_{\rm 2d}+\Gamma^t_{\rm 2d}$, $\Gamma_{\rm TI}\df \Gamma^0_{\rm TI}+\Gamma^t_{\rm TI}$.

\begin{figure}[t]
	\centering
	\includegraphics[width=8.5cm]{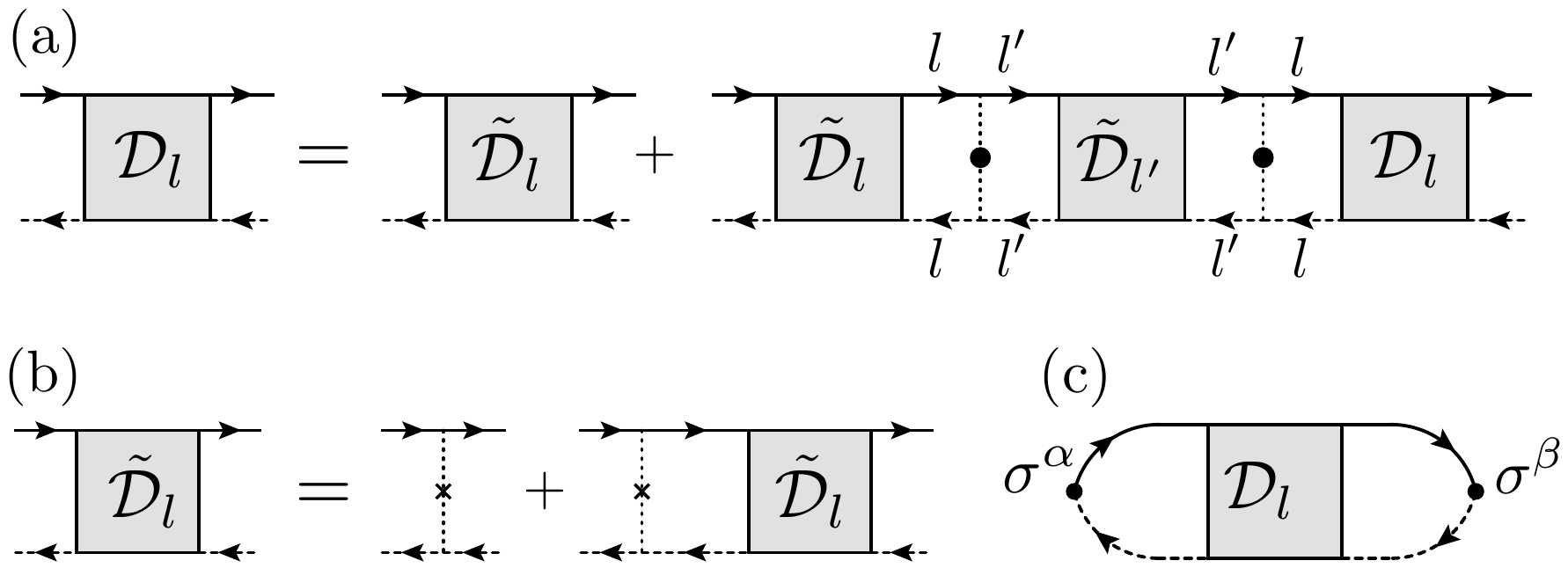}
	\caption{(a) Illustration of the self-consistency equation for the diffuson, Eq.~\eqref{eqn:D2d}. Solid and dashed lines symbolize retarded and advanced Green's functions, respectively. Dotted lines with black circles represent tunneling processes. (b) The Bethe-Salpeter equation for the auxiliary diffuson, Eq.~\eqref{eq:aux}. The dotted lines with crosses represent disorder scattering. (c) The diffuson $\mathcal{D}_l$ can be used to calculate the dynamical part of the response function $\chi_l^{\rm dyn}$ as shown.
	}
	\label{fig:diffuson}
\end{figure}

In the diffusive regime, to leading order in $1/(\eps_F\tau)$, the retarded dynamical part of the spin-density response function for layer $l$, $\chi_{l}^{\rm dyn}$ is obtained by summing all ladder vertex corrections to the bare spin-density response. In our case we have two types of ladder diagrams: the ones due to random interlayer tunneling and the ones due to
intralayer disorder. In most experimentally relevant situations we expect the scattering time due to intralayer disorder
to be much smaller than the relaxation time due to the interlayer random tunneling processes. For 
this reason in the remainder we assume  $\Gamma^t\ll \Gamma^0$. The main building block for the calculation of $\chi^{\rm dyn}_{\rm 2d}$ is the diffuson $\mathcal{D}_{\rm 2d}$, which includes both interlayer tunneling and intralayer ladder diagrams. It satisfies the self-consistent equation~\cite{garate2012,velkov2018}
\beq 
\mathcal{D}_{\rm 2d}= \tilde{\mathcal{D}}_{\rm 2d} + \kappa \tilde{\mathcal{D}}_{\rm 2d}\mathcal{J}_{\rm 2d}^{\rm TI} \tilde{ \mathcal{D}}_{\rm TI} \mathcal{J}_{\rm TI}^{\rm 2d}
            \mathcal{D}_{\rm 2d}.
\label{eqn:D2d}
\enq
In this equation, the auxiliary intralayer diffuson for layer $l$, $\tilde{\mathcal{D}}_{l}$ ($l=(\rm{2d},\rm{TI}))$ is obtained by taking into account only intralyer disorder and the junctions $\mathcal{J}$ describe the transition between the layers. The constant $\kappa$ collects disorder-dependent normalizations with $\kappa^{-1}= n^{\rm 2d}_{\rm imp}n^{\rm TI}_{\rm imp} U^2_{\rm 2d} U^2_{\rm TI}$. The self-consistency equation \eqref{eqn:D2d} is shown diagrammatically in Fig.~\ref{fig:diffuson}(a). 

Mathematically, the auxiliary diffuson $\tilde{\mathcal{D}}_{l}$ satisfies the Bethe-Salpeter equation (see Fig.~\ref{fig:diffuson}(b))
\beq
 \tilde{\mathcal{D}}_{l}(\qq,\omega) = n^l_{imp}U^2_l(\sigma_0 \otimes \sigma_0 - P_{l}(\qq,\omega))^{-1}.
 \label{eq:aux}
\enq
Here, the quantum probability $P_l$ is defined as
\begin{align} 
P_{l}(\qq, \omega)& \equiv n^l_{imp}U^2_l \int_{\kk}  G_{l,\epsilon_F}^R(\kk) \otimes G_{l,\epsilon_F-\omega}^A(\kk-\qq).
\end{align}
The junctions $\mathcal{J}_l^{l'}=P_lt^2P_{l'}$ account for the tunneling processes. The expressions of $P_{\rm 2d}$ and $P_{\rm TI}$ are given in the SM~\cite{SM}.

For the purpose of finding $\chi^{\rm dyn}_l$, it is convenient to solve Eq.~\eqref{eqn:D2d} in the spin-charge representation. To this end the diffusons, as well as the junctions, are contracted with the Pauli matrices as $
 \mathcal{D}_{\rm 2d}^{\alpha\beta}=\frac{1}{2}\sigma_{s_1 s_2}^\alpha \mathcal{D}_{\rm 2d}^{s_1 s_2,s_3 s_4}\sigma_{s_3 s_4}^\beta$ where $\alpha,\beta=(0,x,y,z)$ correspond to the charge and $x,y,z$ components of the spin, respectively. With the knowledge of $\mathcal{D}_l$, the dynamical part of the spin-density response function can be found by introducing charge and spin vertices as illustrated in Fig.~\ref{fig:diffuson}(c). The full response function is then obtained by adding the static part, $\chi_l=\chi_l^{\rm st}+\chi_l^{\rm dyn}$, where $\chi_l^{st,\alpha\beta}\propto \rho_l\delta_{\alpha\beta}$. For systems with conserved particle number, the density response function $\chi^{00}$ must satisfy the condition $\lim_{\omega\rightarrow 0}(\lim_{{\bf q}\rightarrow 0}\chi^{00}({\bf q},\omega))=0$. In the problem under consideration, electrons can move from one layer to the other. Therefore, a complete description of the time evolution of the charge and spin densities must include the mixed response function $\chi_{ll'}$ with $l\ne l'$, i.e. the response of densities in layer $l$ to perturbations in layer $l'$. $\chi_{ll'}$ can be found in analogy to $\chi_l$. 

In the 2DEG, the charge and spin response to external perturbations in the form of electric potentials or Zeeman fields may be conveniently cast in the form of coupled transport equations. In the diffusive limit, we find
\begin{align} 
\partial_t   n_{\rm 2d}   = & \bar{{D}} \nabla^2 \tilde{n}_{\rm 2d} +\Gamma_{ns}l_{\rm TI}(\hat{z}\times \boldsymbol{\nabla})\tilde{\boldsymbol{s}}_{\rm 2d}  -\nu\partial_t(V_{\rm 2d}-V_{\rm TI})
                           \label{eq:diff_1} \\ 
\partial_t {\bf s}_{\rm 2d}=& \left({D}_{\rm 2d}\nabla^2-\Gamma^t_{\rm 2d}\right)\tilde{\bf s}_{\rm 2d}+\alpha\Gamma_{\rm 2d}^t l_{\rm TI}(\hat{z}\times \boldsymbol{\nabla})\times \tilde{\bf s}_{\rm 2d}\nonumber\\
&+\Gamma^t_{\rm 2d}l_{\rm TI}(\hat{z}\times \boldsymbol{\nabla})\left[l_{\rm TI}(\boldsymbol{\nabla}\times \tilde{{\bf s}}_{2d})_z+\tilde{n}_{\rm 2d}/2\right]
	 \label{eq:diff_2} 
\end{align}
where the effective charge diffusion constant
\begin{align}
   \bar{{D}}  = \frac{\Gamma^t_{\rm 2d}D_{\rm TI}+\Gamma^t_{\rm TI}D_{\rm 2d}}{\Gamma^t_{\rm 2d}+\Gamma^t_{\rm TI}},\qquad \Gamma^t_{l}=\frac{1}{\tau^t_{l}},              \label{eq:Ds}
\end{align}
is a weighted average of the diffusion constants ${D}_{\rm 2d}=v_F^2\tau^0_{\rm 2d}/2$, and ${D}_{\rm TI}=v_{\rm TI}^2\tau^0_{\rm TI}$ in the 2DEG and TI, respectively. Moreover, $l_{\rm TI}=v_{\rm TI}\tau^0_{{\rm TI}}$ is the TI mean free path. The spin-charge coupling in the 2DEG is characterized by $\Gamma_{ns}=2\Gamma^t_{\rm 2d}\Gamma^t_{\rm TI}/(\Gamma^t_{\rm 2d}+\Gamma^t_{\rm TI})$. The term containing the dimensionless constant $\alpha={\eps_F\tau^0_{{\rm 2d}}}/(2\pi^2\nu_{\rm TI}D_{\rm TI})$
originates from the induced spin-orbit coupling in the 2DEG.
The charge and spin densities $\tilde{n}$ and $\tilde{\bf s}$ appearing on the right hand side of the diffusion equations include external driving potentials for the charge, $V_{\rm 2d}$, and spin, ${\bf h}_{\rm 2d}$, respectively, as $\tilde{n}_{\rm 2d}=n_{\rm 2d}+2\rho_{\rm 2d}V_{\rm 2d}$ and $\tilde{\bf s}={\bf s}-2\rho_{\rm 2d}{\bf h}_{2d}$. The last term in Eq. (\ref{eq:diff_1}) accounts for a potential loss of electrons in the 2DEG for a dynamically driven system, with coefficient $\nu =2\rho_{\rm 2d} \Gamma^t_{\rm 2d}/(\Gamma^t_{\rm 2d}+\Gamma^t_{\rm TI})$.

Equations~\ceq{eq:diff_1}-\ceq{eq:Ds} are the main result of this work.
They show that in a 2DEG-TI system charge transport and spin transport are coupled 
{\em even} when the tunneling between the two systems is random.
Notice that Eqs.~\ceq{eq:diff_1}-\ceq{eq:Ds} were obtained in the limit in which $\Gamma^t_{l}/\Gamma^0_{l}\ll 1$, and $\omega\tau\ll 1$, $\tau$ being the longest relaxation time: $\tau={\rm max}(\tau^t_{\rm 2d},\tau^t_{\rm TI})$. Eqs.~\ceq{eq:diff_1}-\ceq{eq:Ds}
can only describe transport over time scales much larger than $\tau$ and therefore are not valid in the limit $t=0$ for which $\tau\to\infty$.
For $t=0$ the two systems are decoupled and for the 2DEG the diffusive transport of charge and spin are independent with 
${D}_{\rm 2d}=v_{\rm 2d}^2\tau^0_{\rm 2d}/2$. 

It is instructive to note that there are two mechanisms responsible for the spin-charge and spin-spin coupling in Eqs.~\ceq{eq:diff_1} and \ceq{eq:diff_2}. The term with coefficient $\alpha$ in Eq.~\ceq{eq:diff_2} results from the real part of the self-energy in Eq.~\ceq{eq:sigmat_TI}, i.e. from the tunneling-induced spin-orbit coupling in the effective single-particle Hamiltonian of the 2DEG. This term couples in-plane and out-of-plane spin components. The spin-charge coupling in Eqs.~\ceq{eq:diff_1} and Eqs.~\ceq{eq:diff_2} has a different origin. The surface of the TI hosts a single gapless diffusion mode in the absence of tunneling, as can be seen by diagonalizing the diffuson \cite{burkov2010,garate2012,velkov2018}. For finite ${\bf q}$, this mode has a non-trivial spin structure. By means of the random tunneling, this mode and the gapless modes in the 2DEG hybridize. The hybridization gives rise to spin-charge coupling via the term with coefficient $\Gamma_{ns}$ in Eq.~\ceq{eq:diff_1} and the final term in Eq.~\eqref{eq:diff_2}, as well as to anisotropic spin-diffusion encoded in the first term of the second line in Eq.~\eqref{eq:diff_2}. To leading order in tunneling, the two described mechanisms for spin-charge coupling are independent of each other. As follows from Ref.~\cite{burkov2004}, spin-orbit coupling eventually also leads to spin-charge coupled transport at higher orders in the coupling strength. A separate consequence of the tunneling in Eq.~\eqref{eq:diff_2} is that, since spin is not conserved in the coupled system, a gap of size $\Gamma_{\rm 2d}^t$ opens for the spin diffusion modes.

Equations~\ceq{eq:diff_1},~\ceq{eq:diff_2} show that the strength of the coupling between charge transport and spin transport, 
and the spin-diffusion anisotropy, are proportional
to the ratio $\Gamma^t_{2d}/\Gamma^0_{2d}$. Given that $\Gamma^t_{2d}=t^2\rho_{TI}\pi$, and that $\rho_{TI}$ scales
linearly with $\mu_{TI}$,  we see that both in the 2DEG both the
spin-charge coupling and the spin-diffusion anisotropy
can be tuned simply by changing the doping of the TI's surface. We now study the solution of Eqs.~\ceq{eq:diff_1},~\ceq{eq:diff_2} for a simple setup, as in Refs.~\onlinecite{burkov2004,burkov2010},
to highlight some of the transports effects 
due to the coupling between spin and charge transport described by Eqs.~\ceq{eq:diff_1},~\ceq{eq:diff_2},
and to highlight some of the main similarities and differences between
a 2DEG-TI system, a TI's surface, and a 2DEG with Rashba SOC. We consider a system of size $L$ along $x$, $-L/2<x<L/2$, and in which all the quantities
are uniform along $y$. In the stationary limit, due to the uniformity along $y$,
Eqs.~\ceq{eq:diff_1},~\ceq{eq:diff_2} separate in two independent sets of equations:
one set describing the coupled transport of $n$ and $s^y$, one set describing the coupled
transport of $s^x$ and $s^z$. Given that we are interested
in the coupling between charge and spin transport, we focus on the first set. Due to the assumption that all the quantities are homogenous along $y$,
the coupled equations for $n$ and $s^y$ 
for a 2DEG-TI, a TI, and a 2DEG with Rashba SOC have the same structure:
\begin{align}
    &D_n \partial^2_x n + 2\beta_s \partial_x s^y = 0;                   \label{eq:dn-stat}\\
    &D_s \partial^2_x s^y - \frac{s^y}{\tau_s} + \beta_n \partial_x n = 0. \label{eq:ds-stat}
\end{align}
where $D_n$, $D_s$, $\beta_n$, and $\beta_s$ are constants whose expression in terms of
the parameters characterizing the system are given in Table~\ref{tab:coeff} for a 2DEG-TI, a TI, and a 2DEG with Rashba SOC. From charge conservation, using Eq.~\ceq{eq:diff_1}, we find that the charge current takes the form
$
 \JJ = - \bar D \nabla n_{\rm 2d} -
                          \Gamma_{ns}l_{\rm TI}(s^x_{2d}\hat y -s^y_{2d}\hat x),
$
and for the simple case described by Eq.~\ceq{eq:dn-stat},
$\JJ=J \hat x$, $J = -D_n dn/dx + 2\beta_s s^y$, with $D_n$ and $\beta_s$ given in Table~\ref{tab:coeff}. Similarly from  Eq.~\ceq{eq:ds-stat} we can obtain an expression for the current of $s^y$.
This expression has the term  $\beta_n \partial_x n$, however,
as pointed out before~\cite{rashba2003,malshukov2005,galitski2006,bleibaum2006,tserkovnyak2007},
such term describes an equilibrium spin current and therefore should not be included in the definition of an externally driven spin current.
Knowing the expression of $\JJ$ and of the spin current allows us to write the boundary conditions for
Eqs.~\ceq{eq:dn-stat},~\ceq{eq:ds-stat}, corresponding to the situation when
a charge current $I$ is injected at $x=-L/2$ via
a ferromagnetic electrode so that the incoming electrons have a net spin polarization $\phi$
along $s^y$:
\beq
 \left. J  \right\rvert_{x=\pm \frac{L}{2}} =\frac{I}{e}, \hspace{0.11cm}
 \left. D_s  \partial_x s^y  \right\rvert_{x=-\frac{L}{2}} = -\frac{I \phi}{e},\hspace{0.11cm}
 \left. D_s   \partial_x s^y  \right\rvert_{x=\frac{L}{2}}  = 0,
 \label{eq:BC}
\enq

Recalling that the voltage drop $\Delta V(x)$ \cite{Voltage} at position $x$ is given by $\Delta V(x)=-(1/2e\rho)\int_{-L/2}^x dx'(dn/dx')$,
and solving Eqs.~\ceq{eq:dn-stat},~\ceq{eq:ds-stat} with the boundary conditions~\ceq{eq:BC}
we find
\begin{align}
    s^y(x) & = \frac{I \phi l_{*}}{e D_s } \frac{\cosh{( (x-L/2)/l_{*} )}}{\sinh{(L/l_{*})}} - \frac{l^2_{*} \beta_n I }{e D_n D_s} \label{eq:solSy} 
\end{align}
and the voltage drop between the leads
\begin{equation}
    \Delta V = \frac{I}{2e^2 \rho D_n}  \left(  \frac{2 l^2_{*} \beta_s }{  D_s} \left[\phi - \frac{\beta_n L}{D_n}  \right]  + L \right).
    \label{eq:V}
\end{equation}
In Eqs.~\ceq{eq:solSy},~\ceq{eq:V} $l^{-2}_{*} \equiv  1/(\tau_s D_s) + 2 \beta_n \beta_s/(D_n D_s) $. Using the expressions given in Table~\ref{tab:coeff} for $D_n$, $D_s$, $\tau_s$, $\beta_n$, and $\beta_s$,
Eqs.~\ceq{eq:solSy} and~\ceq{eq:V} for a 2D-TI system become, to leading order in the tunneling amplitude (with $l_*\approx \sqrt{D_{\rm 2d} \tau^t_{2d}}$)
\begin{align}
s^y(x) =&  \frac{I\phi l_*}{eD_{\rm 2d}}
               \frac{\cosh{\left[ (x -L/2)/l_* \right]}}{\sinh{\left( L/l_* \right)}}
            -  \frac{I l_{\rm TI}}{2e\bar{D}} 
               \label{eq:solSy2DTI},\\
\Delta V      =&  \frac{I}{2e^2 \rho_{\rm 2d} \bar{D}}\left(L+2l_{\rm TI}\phi\frac{\Gamma^t_{\rm TI}}{\Gamma^t_{\rm 2d}+\Gamma^t_{\rm TI}}\right).
               \label{eq:solV2DTI}
\end{align}

The second term on the r.h.s. of Eq.~\ceq{eq:solSy2DTI} shows that, as in the case of 2DEG with Rashba SOC~\cite{burkov2004} and a TI~\cite{burkov2010}, 
an Edelstein~\cite{edelstein1990} effect is present, i.e.,
a constant nonequilibrium spin polarization generated by a charge current $I$.
This effect is present due to the ``mirroring'' into the 2DEG of the TI's gapless diffusion mode characterized by the coupling of charge
and spin. It is interesting to notice that for a 2DEG-TI system such term,
as long as $\tau^t_{\rm 2d}\gg \tau^0_{\rm 2d}$ to remain in the regime of validity of the diffusion equations~\ceq{eq:diff_1},~\ceq{eq:diff_2},
is independent of the interlayer tunneling strength. This is due to the fact that in the 2DEG-TI van der Waals structure,
in the 2DEG layer, both the spin relaxation rate, $1/\tau_s$, and the spin-charge coupling $\beta_n$ in Eq.~\ceq{eq:ds-stat}
scale as $t^2$. As a consequence we expect that even in the limit of very small $t$ a significant Edelstein effect should 
be present in a metallic 2D layer placed in proximity of a system with significant SOC such as a TI's surface.
In addition, 
we see that for a 2DEG-TI system, contrary to a TI, 
the strength of the Edelstein effect can be tuned by varying the doping, and therefore $\rho_{\rm TI}$, of the TI's surface. The other important result is that the decay length of $s^y$ is $l^*$ that can also be tuned by varying the doping in the TI,
and that can be very long in the weak tunneling regime,
for wich $\tau^t_{\rm 2d }\gg \tau^0_{\rm 2d }$.
The last term on the r.h.s. of Eq.~\ceq{eq:solV2DTI} is a magnetoresistance contribution to the voltage drop due to the
coupling of the charge and spin transport. For a 2DEG-TI system this term is therefore dependent on the relative strength of the disorder in the TI and 2DEG.

\begin{table}[t]
\centering
\begin{tabular}{ |c || c | c | c |  }
 \hline
   & 2D+TI  & TI & Rashba \\
 \hline
 $D_n$            &  $\bar D$  &  $v^2_{\rm TI} \tau^0_{\rm TI}/2$ & $v^2_{\rm R} \tau^0_{\rm R}/2$ \\
 $D_s$            &  $D_{\rm 2d} + \Gamma^t_{\rm 2d} l^2_{\rm TI}$ & $ 3 D_n/2$  & $D_n$    \\
 $\beta_n$ &  $ \frac{1}{2}\Gamma^t_{\rm 2d} l_{\rm TI}$ & $v_{\rm TI}/2$ & $ -\lambda (\lambda k_F \tau^0_{ \rm R})^2 $ \\
 $\beta_s$ & $ \frac{1}{2}\Gamma_{ns} l_{\rm TI} $&  $v_{\rm TI}/2$  & $2\beta_n$ \\
  $\tau_s$       & $ \tau^t_{2d}$ &  $\tau^0_{TI}$  & $2 \tau^0_{R}/(2\lambda k_F \tau^0_{R})^2$ \\
 \hline 
\end{tabular}
 \caption{Diffusion coefficients for a TI, Rashba 2DEG, and 2D+TI. $\lambda$ is the SOC strength in the Rashba 2DEG, and $\tau^0_{R}$ the Rashba scattering time.
          }
 \label{tab:coeff}
\end{table}

In conclusion, we have studied the electron and spin transport in a van der Waals system formed by one layer with strong spin-orbit coupling and a second layer without spin-orbit coupling, in the regime when the interlayer tunneling is random,
and shown  that in the layer without intrinsic spin-orbit coupling spin-charge coupled transport can be induced 
by the hybridization of the diffusion modes of the two isolated layers.
To exemplify the mechanism we have studied a van der Waals system formed by a 2DEG and TI's surface
and shown how the coupling of the spin and charge transport in the TI is ``mirrored'' into the 2DEG.
In addition,  for the specific case of a 2DEG-TI van der Waals system,
we show that a spin-orbit coupling term is induced into the 2DEG,
and that the induced
coupling of spin and charge transport in the 2DEG can be tuned by varying the TI's doping.
Finally we showed how the coupled spin-charge transport described by the diffusive equations that we obtain for the 2DEG
leads to a current-induced non-equilibrium spin accumulation and 
a magnetoresistance effect that are also tunable by changing the TI's doping.

We thank Ion Garate for useful discussions. This work was supported in part by the US-Israel Binational Science Foundation grants No. 2014345 (M.R.V.), the NSF CAREER grant DMR-1350663 (M.R.V.), the NSF Materials Research Science and Engineering Center Grant No. DMR-1720595 (M.R.V.), the College of
Arts and Sciences at the University of Alabama (G.S.), and
the National Science Foundation under Grant No.
DMR-1742752 (G.S.).
ER acknowledges support from NSF CAREER grant No. DMR-1455233, 
ONR grant No. N00014-16-1-3158, and ARO grant No. W911NF-18-1-0290.
E.R. thanks the Aspen Center for Physics, which is supported by National Science Foundation grant PHY-1607611, for its hospitality while part of this work was performed.


%

\newpage

\begin{center}
 {\bf SUPPLEMENTAL MATERIAL}
\end{center}

\appendix

\section{Expression of  quantum probabilities}

Here, it is convenient to define $\tau_{l}=1/\Gamma_{l}$ for $\Gamma_l=\Gamma^0_l+\Gamma^t_l$ and to display formulas for $\tilde{P}_l=({\tau_{l}^0}/{\tau_{l}})P_l$. In the limit $\Gamma^t_{l}/\Gamma^0_{l}\ll 1$ and $\Gamma^0_{l}/\epsilon_F\ll 1$, 
to leading order in $\omega/\Gamma^0_{l}$ and $v_F q/\Gamma^0_{l}$ we find:
\begin{align} \nonumber
\tilde{P}_{\rm 2d}(\qq, \omega) = & \tilde{a}_{\rm 2d}(\qq, \omega ) \sigma^0 \otimes \sigma^0 + \tilde{b}^a_{\rm 2d}(\qq, \omega)\sigma^0 \otimes \sigma^a \\
& + \tilde{c}^a_{\rm 2d}(\qq, \omega)  \sigma^a \otimes \sigma^0, 
\label{eq:P2d} \\ \nonumber
\tilde{P}_{\rm TI}(\qq, \omega) 
  = & \tilde{a}_{\rm TI}(\qq, \omega) \sigma^0 \otimes \sigma^0 + \tilde{b}_{\rm TI}^a(\qq,\omega) \left( \sigma^0 \otimes \sigma^a + \right. \\
    &\left. \sigma^a \otimes \sigma^0 \right) + \tilde{d}_{\rm TI}^{ab}(\qq, \omega) \sigma^a \otimes \sigma^b\;.
\label{eqn:PTI}
\end{align} 
For the 2DEG:
\begin{align}
\tilde{a}_{\rm 2d}(\qq, \omega) & \approx 1 + i \omega \tau_{\rm  2d} - \tau_{ \rm 2d} \tilde{D}_{\rm 2d } q^2 \\
\tilde{b}^x_{\rm 2d}(\qq, \omega) & \approx \alpha\tau_{\rm 2d}\Gamma^t_{2d}\tilde{l}_{\rm TI} q_y/4=-\tilde{c}^x_{\rm 2d}({\bf q},\omega)   \\
\tilde{b}^y_{\rm 2d}(\qq, \omega) & \approx -\alpha\tau_{\rm 2d}\Gamma^t_{2d}\tilde{l}_{\rm TI} q_x/4=-\tilde{c}^y_{\rm 2d}({\bf q},\omega)
\end{align}
where $\tilde{D}_{ 2d } = v_{2d}^2\tau_{  2d}/2 $ and $\tilde{l}_{\rm 2d}=v_{\rm 2d}\tau_{\rm 2d}$.

For the TI's surface~\cite{burkov2010}
\begin{align} 
\tilde{a}_{\rm TI} &=\left(1- \tau_{{\rm TI}} \tilde{D}_{{\rm TI}} q^2+ i \omega \tau_{{\rm TI}}\right)/2,\\ 
\tilde{b}^x_{\rm TI}&=-i\tilde{l}_{\rm TI} q_y  /4,\quad b^y_{\rm TI}=i\tilde{l}_{\rm TI} q_x  /4,\\ 
\tilde{d}^{xx}_{\rm TI}&=\left(1 -  \tau_{{ TI}}\tilde{D}_{{\rm TI}} (q_x^2 + 3 q_y^2)/2 + i \omega  \tau_{{\rm TI}}\right)/4,\\ 
\tilde{d}^{yy}_{\rm TI}&= \left(1-  \tau_{{\rm TI}}\tilde{D}_{{ \rm TI}} (3q_x^2 +  q_y^2)/2 + i \omega  \tau_{{\rm TI}}\right)/4,\\ 
d^{xy}_{\rm TI}&= d^{yx}_{\rm TI} = {\tau}_{\rm TI} \tilde{D}_{\rm TI}   q_x q_y/4,
\end{align}
where 
$\tilde{D}_{TI} =  v^2_{ TI}\tau_{\rm TI}/2$ and $\tilde{l}_{\rm TI}=v_{\rm TI}\tau_{\rm TI}$. 

\section{Spin-charge diffusion equation for TI's surface}

To facilitate the comparison between the results that we obtain in the main text for a 2DEG-TI system and an isolated TI's surface
we report here the diffusion equations for a TI's surface, first derived in Ref.~\onlinecite{burkov2010}:
\begin{align} 
\label{eq:ti_diff_eqns_a}
\partial_t n_{\rm TI} & = D_{\rm TI} \nabla^2 n_{\rm TI} + v_{\rm TI}(\hat{z}\times \nabla)\cdot \vec{s}_{\rm TI}\\ \nonumber
\label{eq:ti_diff_eqns_b}
\partial_t s^x_{\rm TI} & = \frac{ D_{\rm TI}}{2} \partial^2_x s^x_{\rm TI}+\frac{3  D_{\rm TI}}{2} \partial^2_y s^x_{\rm TI} - D_{\rm TI} \partial^2_{xy} s^y_{\rm TI} \\ & -\frac{s^x_{\rm  TI}}{\tau^0_{\rm  TI}}- \frac{v_{\rm TI}}{2} \partial_y  n_{\rm TI}\\ \nonumber
\partial_t s^y_{\rm TI} & = \frac{3  D_{\rm TI} }{2} \partial^2_x s^y_{\rm TI} +\frac{ D_{\rm TI}}{2} \partial^2_y s^y_{\rm TI} -  D_{\rm TI} \partial^2_{xy} s^x_{\rm TI} \\&-\frac{s^y_{\rm  TI}}{\tau^0_{\rm TI}}+ \frac{v_{\rm TI}}{2} \partial_x n_{\rm TI},
\label{eq:ti_diff_eqns}
\end{align}
where $n_{\rm TI}$ is the carrier density on the TI's surface, and $\vec{s}_{\rm TI}=(s^x_{\rm TI},s^y_{\rm TI})$. Notice that the spin densities are damped by scattering with non-magnetic impurities due to spin-orbit coupling.
Due to a typo in Ref.~\onlinecite{burkov2010} the terms with mix derivatives 
have opposite sign compared to Eqs.~\ceq{eq:ti_diff_eqns_b},~\ceq{eq:ti_diff_eqns_b}.
We can see that the negative sign in front of the terms 
$\partial^2_{xy} s^x_{\rm TI}$, $\partial^2_{xy} s^y_{\rm TI}$
in Eqs.~\ceq{eq:ti_diff_eqns_b},~\ceq{eq:ti_diff_eqns_b}
is correct by considering that when $n_{\rm TI}$ is uniform in time and space
so that Eq.~(\ref{eq:ti_diff_eqns_a}) implies $\partial_y s_{\rm TI}^x = \partial_x s_{\rm TI}^y$,
Eqs.~(\ref{eq:ti_diff_eqns_b}) and (\ref{eq:ti_diff_eqns}) lead to 
$\partial_t s^\alpha_{\rm TI}  = ((1/2)D_{\rm TI} \nabla^2 - 1/\tau^0_{\rm  TI} ) s^\alpha_{\rm TI}$,
the expected spin-diffusion equation in this simple limit.

\section{Diffusion equations for two coupled 2DEGs}
In this appendix, we review the density diffusion equation of a 2DEG-2DEG heterostructure. The effect of the coupling in the quantum interference has been studied before \cite{bergmann1989}. Each layer $l$ posses its own diffusion constant $D_{l}$ and density of states $\rho_{l}$, where $l=T,B$ labels the top and bottom 2DEG layer respectively. We obtain
\begin{align} \nonumber
\partial_t n^{(T)}_{\rm 2d} & = \bar D \nabla^2 n^{(T)}_{\rm 2d},
\label{eq:2d-2d_density}
\end{align}
where we have defined
\begin{equation}
    \bar D = \frac{\Gamma^t_T D_B + \Gamma^t_B D_T }{\Gamma^t_T + \Gamma^t_B}, \;\;\; \Gamma^t_l = \frac{1}{\tau^t_l}
\end{equation}
The renormalized diffusion constant contains corrections proportional to the diffusion constant in the bottom layer. The leading corrections to the diffusion constant is given by a term proportional the ratio of the DOS in each layer. Given that there is no spin-orbit coupling, the spin follow analogous diffusion equations in each direction. 


\begin{thebibliography}{42}
\bibitem [{Vol()}]{Voltage}%
  \BibitemOpen
  \href@noop {} {}\bibinfo {note} {Here we make the electric charge explicit,
  $V_n=eV$.}\BibitemShut {Stop}%
\end{thebibliography}
\end{document}